# Comparison of polarization and Born effective charges in alternatively-deflected zigzag and planar-zigzag $\beta$-poly(vinylidene fluoride)

*Nicholas J. Ramer\* and Kimberly A. Stiso*

*Department of Chemistry, Long Island University – C. W. Post Campus,*

*Brookville, New York 11548-1300*

## ABSTRACT

A density-functional study of the alternatively-deflected zigzag structure for the β-phase of poly(vinylidene fluoride) or $(–CH_2–CF_2–)_n$ is presented. Structure, polarization and effective charges have been determined. Using lattice constants and atomic positions transformed from experiment, the space group for the alternatively-deflected zigzag structure was found to be *Cc*2*m* ($C_{2v}^{16}$), in agreement with previous studies. We find a polarization for the alternatively-deflected structure identical to that of the planar-zigzag structure with small changes in the accompanying effective charges.

PACS numbers: 77.84.Jd, 61.66.Hq, 31.15.Ew

\* Corresponding author.

1. INTRODUCTION

The piezoelectric and pyroelectric properties of poly(vinylidene fluoride) or PVDF have been known for over thirty years.[1,2] The ferroelectric nature of the polarization was first reported by Furukawa *et al*.[3] The physical properties of PVDF, specifically its flexibility, low density, low mechanical impedance and ease of fabrication have made it and its related co-polymers very attractive for device applications.

The monomeric repeat unit of PVDF is ($-CH_2-CF_2-$). There are four known crystalline forms of PVDF: $\alpha$, $\beta$, $\gamma$ and $\delta$ phases.[4] The phases differ in crystallographic space group, number of formula units per unit cell, packing of polymer chains and patterning of *trans* and *gauche* conformations of the $-CH_2-$ and $-CF_2-$ groups. The $\beta$- and $\delta$-phases exhibit ferroelectricity with the $\beta$-phase being studied more extensively both theoretically and experimentally.

Two structures have been proposed in the literature for $\beta$-PVDF; an all-*trans* planar-zigzag form (see Figure 1(a)) (Refs.5,6,7,8) and an alternatively-deflected zigzag form (see Figure 1(b)).[5,9] These two slightly different structure have been proposed based on a consideration of van der Waals radii and fluorine-fluorine distance. The use of the term zigzag refers to the conformation of carbon-carbon bonds in the polymer backbone. Planar-zigzag indicates that all carbons in the chain lie in a plane (in this case the *yz* plane) while an alternatively-deflected zigzag structure has its carbons attached to the hydrogens all lying in the *yz* plane with the carbons attached to the fluorines alternating in deflection direction ($\pm x$) from the *yz* plane. In addition, the alternatively-deflected structure is reported to possess $-CH_2-$ rotations within the *xy* plane.[9] In either structure, a spontaneous polarization is generated parallel to the *y*-axis.

Before describing the origin of an alternatively-deflected zigzag structure, it is important to summarize the clearest experimental evidence supporting the planar-zigzag structure. In the planar-zigzag structure, the fluorine-fluorine distance is reported as 2.56 Å, which is equal to the $c$ lattice constant of the reported orthorhombic unit cell. This distance is 0.14 Å less than twice the van der Waals radius of fluorine ($2 \times 1.35$ Å $= 2.70$ Å). Lando *et al*. propose a crowding of the fluorine atoms along the chain axis and that atomic motion will be restricted along this axis.[8] This conclusion is supported by their observation of the (00$l$) X-ray reflections having a temperature factor, $B$, equal to 3.8 Å$^2$. This $B$ factor is considerably less than that for other observed reflections in the diffraction pattern having non-zero $h$ and $k$ indices ($B$=7.0 Å$^2$). In addition, Lando *et al*. determined the NMR second moment as a function of the alignment angle between the magnetic field and the sample draw direction. The minimum second moment was found at an alignment angle of 45º indicating the probability of a planar-zigzag confirmation.

The crowding of the fluorine atoms was also recognized by Galperin *et al*. and they tentatively proposed a structure with –CF$_2$– deflections, with neighboring –CF$_2$– groups deflected in opposite directions.[5] These deflections would accomplish enlarging the fluorine-fluorine distance. They suggest a deflection angle of approximately 5º. To confirm the existence of a deflected structure, Hasegawa *et al*. determined the discrepancy factor, $R$,

$$R(\sigma) = \frac{\sum |F_{obs} - F_{calc}(\sigma)|}{\sum |F_{obs}(\sigma)|}$$

as a function of deflection angle, $\sigma$, for their observed X-ray data where $F_{obs}$ and $F_{calc}$ are the observed and calculated structure factors.[9,10] For their planar-zigzag structure ($\sigma$=0º), Hasegawa *et al*. determined an $R$ of 18.5%. By varying $\sigma$, they determined a minimum $R$ of 13.5% for σ=7°.[10] At this deflection angle, the fluorine-fluorine distance increases to 2.60 Å. This result supported their identification of an alternatively-deflected zigzag structure for β-PVDF.

However, as correctly indicated by Lando *et al.*, these deflections would be accompanied by a doubling of the repeat distance along the chain axis (2 × 2.56 Å = 5.12 Å).[8] Experimentally, an intermediate layer line with that period is not present in the diffraction pattern.[9] Furthermore, the presence of these deflections would mean that the assigned space group of *Cm*2*m* would no longer be correct.

In response to these contradictions for the existence of an alternatively-deflected structure, Hasegawa *et al.* propose a statistically-deflected zigzag structure.[9] In this model, two alternatively-deflected zigzag chains are present in β-PVDF, each chain having equal probability within the lattice. The statistically-disordered model therefore restores the observed periodicity, chain axis length and space group. Additional experimental evidence for the presence of a deflected structure can be found in an analysis of β-PVDF vibrational spectra.[11] Specifically, several modes with weak absorption in the far-infrared region are attributed to the presence of a deflected chain model based upon analysis of the dispersion relations and frequency distribution functions.

There are few theoretical studies comparing the two proposed structures of β-PVDF. Elastic constants have been computed using a point-charge model.[12] Young's moduli ($Y$) were determined for both planar-zigzag and alternatively-deflected structures. Only a small change in $Y$ in the chain direction (~15 GPa) was found for the deflected structure as compared to the planar form. Based upon this work, it appears that the deflections have little effect on the mechanical properties for the polymer. A recent study by Duan *et al.* investigated the alternatively-deflected structure employing full-potential linear-augmented plane-wave method (FLAPW).[13] Energy gap and electronic band structure symmetries were determined and shown to be in good agreement with experiment.

This study will ascertain whether there exist any differences in the polarization properties due to the –$CF_2$– deflections in the alternatively-deflected zigzag structure of β-PVDF as compared to the planar-zigzag form. A study of the polarization properties of the planar-zigzag form has been submitted elsewhere.[14] More importantly for future theoretical studies, we will be able to quantify the differences in the polarization properties between the two structures. We will begin by determining the space group and relaxed atomic positions for the alternatively-deflected zigzag structure. Using the relaxed structure, we will compute the spontaneous polarization and the corresponding Born dynamic effective charges.

## 2. METHODS

We have applied density-functional theory within the generalized-gradient approximation (GGA) (Ref. 15) and optimized pseudopotential generation methods were used.[16] For the carbon pseudopotential, a $2s^2 2p^2$ reference configuration was chosen with $r_c(s) = 0.84$ a.u. and $r_c(p) = 1.29$ a.u. The fluorine pseudopotential was constructed from a $2s^2 2p^5$ reference state with $r_c(s) = 1.14$ a.u. and $r_c(p) = 1.63$ a.u. The reference configuration for the hydrogen pseudopotential was $1s^1$ with $r_c(s) = 0.77$ a.u. A plane-wave cut-off energy of 50 Ry was used.

As mentioned above, in order to resolve the observed space group for the β-PVDF sample (*Cm*2*m*) and the existence of a deflected chain, Hasegawa *et al.* propose a statistically-disordered structure with the deflected chain and its mirror image present with equal probability in the lattice.[9] In the present study we do not include this feature of their model. Instead, the unit cell contains two formula units (*Z*=2). Each formula unit contains two monomers with the deflections and periodicity shown in Figure 1(b). The lattice constants [*a* = 8.58 Å, *b* = 4.91 Å, and *c* = 5.12 Å] as determined by Hasegawa *et al.* were used.

The atomic positions for two possible space groups (see below) were determined. For each structure, atomic relaxations will be completed using density-functional calculations. Brillouin zone integrations were done using a 4 × 4 × 4 Monkhorst-Pack *k*-point mesh.[17] Regardless of space group, this yields 8 *k*-points in the irreducible wedge of the Brillouin zone. Relaxed atomic positions were found when the computed Hellmann-Feynman forces on all atoms not constrained by symmetry were less than 0.003 eV/Å.[18] For the determination of the polarization properties, a 4 × 8 × 4 Monkhorst-Pack *k*-point mesh was found to give converged spontaneous polarization, $P_s$ and effective charges, $Z^*$. The Berry-phase method was employed to compute the electronic portion of $P_s$.[19,20,21] The ionic portion of $P_s$, $P_{ion}$, is computed as a lattice summation according to

$$P_{ion} = \frac{e}{V} \sum_m Z_m r_m$$

where *V* is the unit cell volume, $Z_m$ is the core charge of atom *m* and $r_m$ is the position of the *m*-th atom in the cell. Since the Berry-phase method determines a polarization difference between two configurations of the system connected by an adiabatic transformation pathway, a non-polar reference state must be chosen. We have chosen a high-symmetry structure which restores the mirror-plane symmetry to the *y*-axis as the reference state.

### 3. RESULTS AND DISCUSSION

#### A. Space group and atomic positions

As correctly described by Lando *et al.*, a deflected chain conformation will not be compatible with the observed space group of *Cm*2*m*.[8] Instead, a space group isomorphic with *Cm*2*m* but with a doubled *c*-axis must be used. There are two such space groups; *Cc*2*m* ($C_{2v}^{16}$) and *Ic*2*m* ($C_{2v}^{22}$).[22] Each space group contains two formula units; one at the origin and the other at either (½,½,0) for *Cc*2*m* and (½,½,½) for *Ic*2*m*. They both possess the same point symmetries as well as the multiplicity of each Wyckoff position. A previous density-functional theory study using full-potential linear-augmented-plane-wave method (Ref. 13) determined the energetics of the β-PVDF structure for the two space groups with atomic coordinates obtained from fitting bond lengths and angles with values from a previous structural optimization study using force-field calculations.[23] The force-fields were parameterized for a planar-zigzag structure. The density-functional study found the structure with the space group *Cc*2*m* was consistently lower in total energy than the structure with the *Ic*2*m* space group for various *k*-point densities.

In the present study, we have re-examined the assignment of the space group for the alternatively-deflected structure beginning with the atomic coordinates found by Hasegawa *et al.* for their statistically-disordered model.[9] Some considerations must be made in order to utilize these coordinates as a starting point for atomic relaxations in the doubled unit cell. First, the *c*-axis length in the X-ray study was reported as 2.56 Å. This was based on the assignment of the space group *Cm*2*m* and the proposal of a statistically-disordered model for the structure. Since our unit cell is doubled along the *c*-axis, the atomic coordinates along this direction must be halved for our initial structure. Secondly, since the statistically-disordered model includes a mirror image of each polymer chain, the reported atomic coordinates at first glance do not appear to satisfy the necessary symmetries for the *Cm*2*m* space group. In order to resolve this, one must consider two analogous unit cells occurring with equal probability throughout the lattice (see Figure 1(c)). One unit cell is the reflection of the other in the *yz* plane. We have chosen not to include the statistical disorder in our calculations; however, we have incorporated the reported chain's mirror image to extend the periodicity of each chain (see Figure 1(b)) to give the alternatively-deflected pattern. In other words, we have translated the reported positions' mirror image by one-half the unit cell in the *z*-direction. Finally, the positions of the hydrogen atoms deserve particular attention. Because of the transformation to the doubled unit cell, the positions should transform to positions with multiplicity of eight (*c* Wyckoff notation). Because of the disordered model, the hydrogen atoms are separated into two groups; H1 and its equivalent positions and H2 and its equivalent positions (both sites with *d* Wyckoff notation). To resolve this, we have allowed the initial positions of the hydrogen atoms to be those of Hasegawa *et al*.[9] Table I shows the transformation of the experimental coordinates to those used as an initial structure for both space groups in the present study. All the above considerations have been incorporated into the determination of the initial atomic coordinates for the density-functional calculations.

We began our minimization using the transformed atomic coordinates of Hasegawa *et al*. (Table I) but have allowed for the inequivalence of the hydrogen atom positions.[9] In order words, we have permitted –CH$_2$– rotation in the *xy* plane, although forbidden by the symmetry of either space group (see Figure 2(a)). By minimization of the Hellmann-Feynman forces using a conjugate gradient algorithm, we have determined relaxed atomic positions for each space group. For both *Cc*2*m* and *Ic*2*m* space groups, the hydrogen atom coordinates relaxed into equivalent positions consistent with *c* Wyckoff assignments in each space group and multiplicity of eight. We have in principle found an alternatively-deflected structure that has averaged –CH$_2$– rotations in the *xy* plane as compared to the statistically-deflected model (see Figure 2(b)). In addition, the structure with space group *Cc*2*m* was 0.018 eV/unit cell lower in energy that the *Ic*2*m* structure. This confirms the same determination of the space group for the alternatively-deflected structure by previous density functional study.[13]

The relaxed atomic coordinates for the *Cc*2*m* space group structure for alternatively-deflected zigzag β-PVDF are contained in Table II. C1 and C2 are the carbons attached to the H1/H2 and F1/F2 atoms respectively. For comparison we have included the experimental coordinates for the statistically-disordered deflected structure determined by Hasegawa *et al*.[9] The method for determining these positions have been described above and are a result of a minimization of the discrepancy factor for the X-ray data. We find good agreement between our theoretical and the experimental structures. Several factors must be considered in the comparison of the experimental and theoretical values. First, the experimental values are determined at a finite temperature above 0 K. This means that some amount of thermal broadening will be present in the assignment of the atomic positions. In contrast, the theoretical positions are considered to be at 0 K. In addition, the positions of lighter atoms (such as H) are more difficult to ascertain.

Finally, the theoretical structure assumes that only –$CF_2$– deflections are present in the structure. The experimental structure assumes –$CH_2$– rotations as well.

For a better assessment of the positions, we have computed bond lengths and bond angles for both the theoretical and experimental structures. Those values are reported in Table III. The experimental values indicate the improved refinement that prompted the proposal of the deflected chain model by Hasegawa *et al*. when compared to the same values in the planar-zigzag structure.[9] We find very good agreement between our bond lengths and angles with those of the X-ray data. The largest discrepancy is in the C–F bond lengths (~ 3%). We attribute all of the variations to the differences in the theoretical and experimental structures, namely the lack of –$CH_2$– rotations and small $\sigma$ found in the theoretical structure.

## B. Spontaneous polarization

Based on the structure found from atomic relaxations, the polarization was computed using a Berry-phase method for alternatively-deflected zigzag β-PVDF. We find a value of 0.181 C/m$^2$ per unit cell volume. This $P_s$ is exactly the same as the polarization found in the planar-zigzag structure of β-PVDF. This result is not surprising considering that the deflections can be considered a rotation of the –$CF_2$– groups in the *xy*-plane along with a displacement in the ±*x* direction. Since the angle of F1–C2–F2 and C–F bond length in the alternatively-deflected zigzag structure remain relatively unchanged as compared to the same values in the planar-zigzag structure, there will be a negligible contribution if any to the macroscopic polarization of the material. This result would indicate that the effective charges for the alternatively-deflected structure should be very similar to those in the planar-zigzag structure.

C. Dynamic Born effective charges

Using the atomic positions in Table II, Born effective charges were determined for the alternatively-deflected β-PVDF structure. Atomic displacements of approximately 0.1% of the *b* lattice constant were made in the direction of the polarization. For each displaced structure, the spontaneous polarization was determined from density-functional calculations. Using the polarization difference between the displaced structure and the un-displaced structure, the Born effective charge ($Z^*$) was determined for each atom. The computed $Z^*$ values for the alternative-deflected zigzag structure are contained in Table IV. Convergence in *k*-point mesh was confirmed by the sum of $Z^*$ for each atom *m*,

$$\sum_{m=1}^{N} Z_m^*$$

where *N* is the total number of atoms within the unit cell, equaling -0.01.

For comparison we have also included the $Z^*$ for the planar-zigzag structure in Table IV.[14] First, the $Z^*$ for the corresponding atoms in each structure are very similar. This underscores the slight structural differences between the two structures. Once again, just as was seen in the planar-zigzag structure, the $Z^*$ for the fluorine atoms are close to their nominal value (-1) and the $Z^*$ for the hydrogen atoms differ considerably from their nominal value (+1). This can be attributed to the relative polarity of these covalent bonds.

The $Z^*$ for the fluorine atoms in the two structures show the greatest differences as should be expected since the deflections in the alternatively-deflected zigzag structure involve the $-CF_2-$ moieties. The difference between $Z^*(F1)$ and $Z^*(F2)$ for the alternatively-deflected structure can also be attributed to the deflections. Stated simply, the dipole associated with C2–F1 bond has a smaller component parallel to the polarization direction due to the bond's rotation in the *xy*-plane as compared to the planar-zigzag structure. This explains the smaller $Z^*$(F1) in the alternative-deflected structure. Likewise, the dipole associated with the C2–F2 bond has a larger component parallel to the polarization direction and results in a larger $Z^*$(F2).

The deflections also do not appear to have a considerable effect on the charge transfer between the –CH$_2$– and –CF$_2$– groups. In the planar-zigzag structure, the C1–C2 bond shows a small degree of charge transfer as evident by the sum of the $Z^*$ for the –CH$_2$– and –CF$_2$– groups (+0.08/-0.08 respectively). These same bonds in the alternatively-deflected structure possess a similar degree of charge transfer with the $Z^*$ sum for the –CH$_2$– of +0.05 and for the –CF$_2$– group of -0.04.

## 4. CONCLUSIONS

In summary, we have determined the atomic structure of the alternatively-deflected structure of the β-phase of poly(vinylidene difluoride). In addition, we have shown that the space group for the alternatively-deflected structure is *Cc*2*m*. Using this structure, we have determined the spontaneous polarization, which is almost identical to that of the planar-zigzag form of β-PVDF. The effective charges for the two structures are also very similar with only small differences in their values. Overall, this indicates that the deflections have little effect on the polarization properties of β-PVDF and gives a good indication that either model can be used in determining electronic or mechanical properties.

## ACKNOWLEDGMENTS

The authors wish to thank Ilya Grinberg for his invaluable assistance with the calculations and for discussions regarding their results. This work was supported by a grant from the Research Committee of the C. W. Post Campus of Long Island University. Funding for computational equipment was provided by Long Island University. Additional computational support was provided by the Department of Chemistry of the University of Pennsylvania.

TABLES

TABLE I. Experimental (space group *Cm*2*m*) and transformed atomic positions for isomorphic space groups *Cc*2*m* and *Ic*2*m*. Positions designated with (*) represent the mirror image present in the statistically-disordered model. See text for description of transformations. All positions are given as fractions of lattice constants.

| Atom | *Cm*2*m*[a] | | | *Cc*2*m*[b,c] or *Ic*2*m*[b,d] | | |
|------|------|------|------|------|------|------|
|      | x/a  | y/b  | z/c  | x/a  | y/b  | z/c  |
| C1   | 0.000 | 0.000 | 0.000 | 0.000 | 0.000 | 0.000 |
|      | 0.000 | 0.000 | 0.000* | 0.000 | 0.000 | 0.500 |
| C2   | 0.012 | 0.173 | 0.500 | 0.012 | 0.173 | 0.250 |
|      | -0.012 | 0.173 | 0.500* | -0.012 | 0.173 | 0.750 |
| F1   | 0.148 | 0.305 | 0.500 | 0.148 | 0.305 | 0.250 |
|      | -0.148 | 0.305 | 0.500* | -0.148 | 0.305 | 0.750 |
| F2   | -0.102 | 0.359 | 0.500 | -0.102 | 0.359 | 0.250 |
|      | 0.102 | 0.359 | 0.500* | 0.102 | 0.359 | 0.750 |
| H1   | 0.094 | -0.149 | 0.000 | 0.094 | -0.149 | 0.000 |
|      | -0.094 | -0.149 | 0.000* | -0.094 | -0.149 | 0.500 |
| H2   | -0.112 | -0.111 | 0.000 | -0.112 | -0.111 | 0.000 |
|      | 0.112 | -0.111 | 0.000* | 0.112 | -0.111 | 0.000 |

[a] Statistically-disordered structure from reference 9 with $a = 8.58$ Å, $b = 4.91$ Å, and $c = 2.56$ Å.

[b] Alternatively-deflected structure with $a = 8.58$ Å, $b = 4.91$ Å, and $c = 5.12$ Å.

[c] Additional positions are obtained by translating coordinates by (½,½,0).

[d] Additional positions are obtained by translating coordinates by (½,½,½).

TABLE II. Theoretical and experimental atomic positions for alternatively-deflected structure of β-poly(vinylidene fluoride). Positions are given as fractions of lattice constants.

|      | Theory[a] | | | Experiment[b] | | |
|------|-------|-------|-------|--------|--------|-------|
| Atom | x/a   | y/b   | z/c   | x/a    | y/b    | z/c   |
| C1   | 0.000 | 0.000 | 0.000 | 0.000  | 0.000  | 0.000 |
| C2   | 0.005 | 0.170 | 0.250 | 0.012  | 0.173  | 0.250 |
| F1   | 0.141 | 0.325 | 0.250 | 0.148  | 0.305  | 0.250 |
| F2   | -0.115 | 0.359 | 0.250 | -0.102 | 0.359  | 0.250 |
| H1   | 0.103 | -0.132 | 0.000 | 0.094  | -0.149 | 0.000 |
| H2   | -0.103 | -0.132 | 0.000 | -0.112 | -0.111 | 0.000 |

[a] $a = 8.58$ Å, $b = 4.91$ Å, and $c = 5.12$ Å.

[b] Statistically-disordered structure from reference 9 with $a = 8.58$ Å, $b = 4.91$ Å, and $c = 5.12$ Å. Since $c$-axis lattice constant is doubled in present study, reported $z$ coordinates have been halved.

TABLE III. Theoretical and experimental bond lengths and bond angles for the alternatively-deflected structure of β-poly(vinylidene fluoride) computed from reported atomic positions. Values in parentheses indicate quantities used in fitting X-ray data to determine experimental atomic positions (See Table II). Bond lengths are given in Angstroms and angles in degrees.

|           | Theory | Experiment[a]  |
|-----------|--------|----------------|
|           | Bond Lengths ||
| C1–H1     | 1.10   | 1.09(1.09)     |
| C1–H2     | 1.10   | 1.10(1.09)     |
| C2–F1     | 1.38   | 1.33(1.34)     |
| C2–F2     | 1.38   | 1.34(1.34)     |
| C1–C2     | 1.53   | 1.54(1.54)     |
|           | Bond Angles ||
| H1–C1–H2  | 107.1  | 108.7(112)     |
| F1–C2–F2  | 105.7  | 108.0(108)     |
| C1–C2–C1  | 113.6  | 111.9(112.5)   |

[a] Statistically-disordered structure from reference 9 with $a = 8.58$ Å, $b = 4.91$ Å, and $c = 5.12$ Å.

TABLE IV. Computed Born effective charges ($Z^*$) for the alternatively-deflected zigzag structure of β-poly(vinylidene fluoride). For comparison, $Z^*$ for the planar-zigzag structure of β-poly(vinylidene fluoride) are also given.

| Atom | $Z^*$ Alternatively-deflected zigzag | Planar-zigzag[a] |
|---|---|---|
| C1 | -0.23 | -0.20 |
| C2 | 1.45 | 1.44 |
| F1 | -0.70 | -0.76 |
| F2 | -0.79 | -0.76 |
| H1 | 0.14 | 0.14 |
| H2 | 0.14 | 0.14 |

[a] Reference 14.

FIGURES

FIG. 1. Molecular structures for β-poly(vinylidene fluoride). (a) Planar-zigzag, (b) alternatively-deflected zigzag, (c) statistically-disordered alternatively-deflected zigzag structures. See text for description of structure.

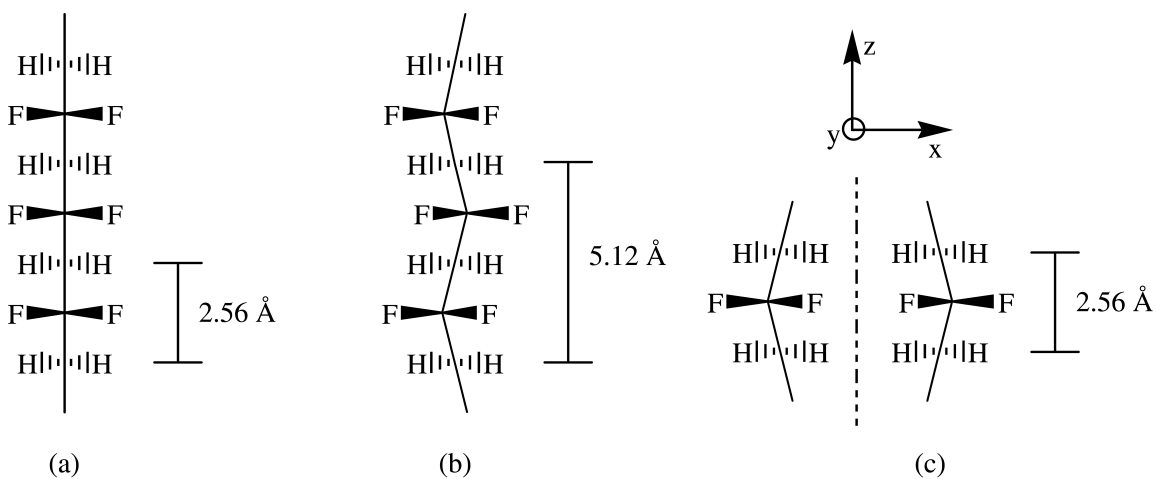

FIG. 2. Chain axis view of β-poly(vinylidene fluoride). (a) Statistically-disordered alternatively-deflected zigzag structure with –CH$_2$– $xy$ rotations according to reference 9, (b) alternatively-deflected zigzag structure with averaged –CH$_2$– $xy$ rotations found by atomic relaxations in the present study. σ is the deflection angle of carbon backbone.

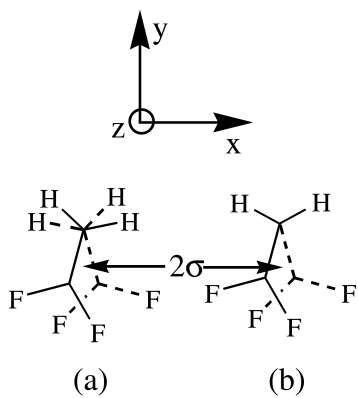

# REFERENCES

[1] H. Kawai, Jpn. J. Appl. Phys. **8**, 975 (1969).

[2] J. G. Bergman, Jr., J. H. McFee, and G. R. Crane, Appl. Phys. Lett. **18**, 203 (1971).

[3] T. Furukawa, M. Date, E. Fukada, Y. Tajitsu and A. Chiba, Jpn. J. Appl. Phys. **19**, L109 (1980).

[4] A. Lovinger, Science **220**, 1115 (1983).

[5] Y. L. Galperin, Y. V. Strogalin, and M. P. Mlenik, Vysolomolekul. Soedin. **7**, 933 (1965).

[6] Y. L. Galperin and B. P. Kosmynin, Vysolomolekul. Soedin. **11**, 1432 (1969).

[7] G. Cortili and G. Zerbi, Spectrochim. Acta **23A**, 2216 (1967).

[8] J. B. Lando, H. G. Olf, and A. Peterlin, J. Polym. Sci. Part A-1 **4**, 941 (1966).

[9] R. Hasegawa, Y. Takahashi, Y. Chatani, and H. Tadokoro, Polym. J. **3**, 600 (1972).

[10] R. Hasegawa, Y. Takahashi, Y. Chatani, and H. Tadokoro, Polym. J. **3**, 591 (1972).

[11] M. Kobayashi, K. Tashiro, and H. Tadokoro, Macromolecules **8**, 158 (1975).

[12] K. Tashiro, M. Koyabashi, H. Tadokoro, and E. Fukada, Macromolecules **13**, 691 (1980).

[13] C.-D. Duan, W. N. Mei, J. R. Hardy, S. Ducharme, J. Choi, and P. A. Dowben, Europhys. Lett. **61**, 81 (2003).

[14] N. J. Ramer and K. A. Stiso (unpublished).

[15] J. P. Perdew, K. Burke, and M. Ernzerhof, Phys. Rev. Lett. **77**, 3865 (1996).

[16] A. M. Rappe, K. M. Rabe, E. Kaxiras, and J. D. Joannopoulos, Phys. Rev. B **41**, R1227 (1990).

[17] H. J. Monkhorst and J. D. Pack, Phys. Rev. B **13**, 5188 (1976).

[18] H. Hellmann, *Einfuhrung in die Quantumchemie* (Deuticke, Leipzig, 1937); R. P. Feynman, Phys. Rev. **56**, 340 (1939).

[19] M. V. Berry, Proc. R. Soc. London, Ser. A **392**, 45 (1984).

[20] R. D. King-Smith and D. Vanderbilt, Phys. Rev. B **47**, R1651 (1993).

[21] R. Resta, Rev. Mod. Phys. **66**, 899 (1994).

[22] *International Tables for Crystallography*, Vols. I, II, III, Kynoch Press, Birmingham England, 1962.

[23] N. Karasawa and W. A. Goddard III, Macromolecules **25**, 7268 (1992).

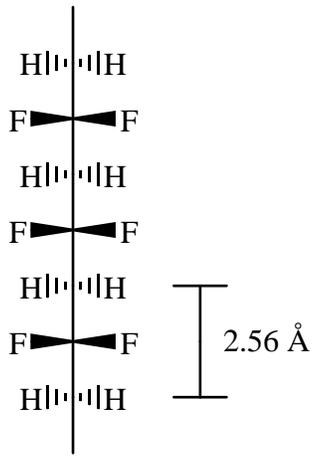 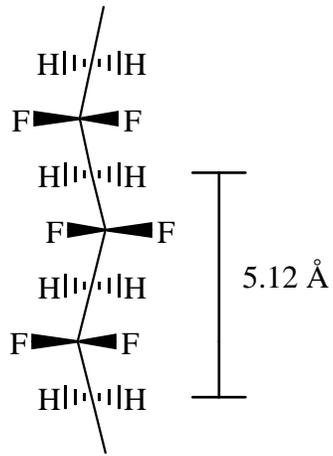 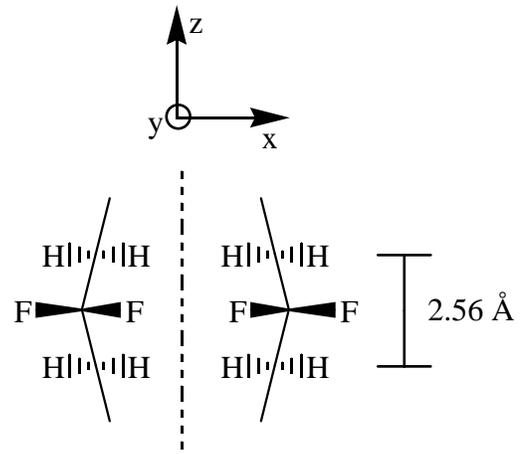

(a)  (b)  (c)

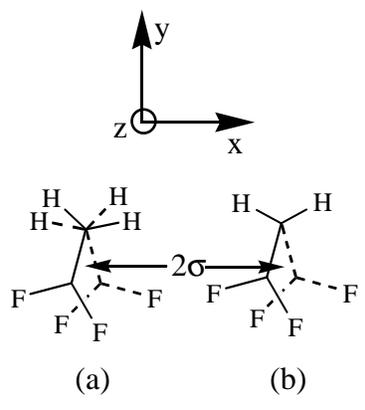

(a) (b)